\documentstyle[12pt]{article}
\def\l{\lambda}

\def\ket#1{|{#1}\rangle}  
\def\bra#1{\langle{#1}|}  
\def\norm#1#2{\langle{#1}|{#2}\rangle}   

\def\cir#1{~~~\begin{picture}(5,5)
\circle{5}
\put(-1,-1){#1}
\put(-5,4.5){\vector(1,-1){3}}
\put(5,4.5){\vector(-1,-1){3}}
\put(-0.3,1.1){\vector(-1,-1){3}}
\put(0.3,1.1){\vector(1,-1){3}}~
\end{picture}}
\def\cirr#1{~~~\begin{picture}(5,5)
\circle{6}
\put(-3,-1){#1}
\put(-5,4.5){\vector(1,-1){2.5}}
\put(5,4.5){\vector(-1,-1){2.5}}
\put(-1.4,0.5){\vector(-1,-1){2.5}}
\put(1.4,0.5){\vector(1,-1){2.5}}~
\end{picture}}
\begin{document}
\begin{titlepage}
\rightline{IMSc/94/45,~~ November 94.}
\baselineskip=24pt
\begin{center}
{\large\bf Representations of Composite Braids and Invariants
for Mutant Knots and Links in Chern-Simons Field Theories}

  \vspace{.5cm}

   {\bf P. Ramadevi, T.R. Govindarajan and R.K. Kaul}\\
   The Institute of Mathematical Sciences, \\
   Taramani, Madras-600 113, INDIA.
\end{center}
\noindent{\bf Abstract}

We  show that any of the new knot invariants obtained from Chern-Simons
theory based on an arbitrary non-abelian gauge group do not distinguish
isotopically inequivalent mutant knots and links. In an attempt to
distinguish these knots and links, we study Murakami (symmetrized version)
$r$-strand composite braids. Salient features of the theory of such composite
braids are presented. Representations of generators for these braids are
obtained by exploiting properties of Hilbert spaces associated with the
correlators of Wess-Zumino conformal field theories. The $r$-composite
invariants for the knots are given by the sum of elementary Chern-Simons
invariants associated with the irreducible representations in the product
of $r$ representations (allowed by the
fusion rules of the corresponding Wess-Zumino conformal field theory)
placed on the $r$ individual strands of the composite braid. On
the other hand, composite invariants for links are given by a weighted sum
of elementary multicoloured Chern-Simons invariants.
Some mutant links can be
distinguished through the composite invariants, but
 mutant knots do not share this property.
The results, though
developed in detail within the framework of $SU(2)$ Chern-Simons theory
are valid for any other non-abelian gauge group.
\vskip1cm
\hrule
\vskip.3cm

\noindent E-mail : rama, trg, kaul@imsc.ernet.in

\end{titlepage}

\section{Introduction}

Since the original work of Jones \cite{jon}, there has
been a lot of interest in polynomial
invariants associated with knots \cite {oce}-\cite {witt}.
Following Witten's pioneering work,  numerous new knot
invariants have been obtained from Chern-Simons theory  based on a
compact semi-simple gauge group \cite {witt}-\cite {kau}.
The expectation values of the Wilson loops, which are the observables
of the Chern-Simons theory, give these new knot invariants.
An alternative method of obtaining these invariants involves study
of $N$-state vertex models \cite {akut}.
Representation theory of  quantum groups provides yet another
framework in which these invariants can be studied \cite {Res}.

Two of the outstanding problems of knot theory are (i) detection of
chirality of knots and links and (ii) distinguishing isotopically
distinct mutant knots and links through the polynomial invariants.
It is well known that Jones, HOMFLY and
Kauffman/Akutsu-Wadati polynomials do not detect chirality of some of the
knots.
Within Chern-Simons field theoretic framework, Jones and Akutsu-Wadati
polynomial correspond to $SU(2)$ theory with spin $1/2$ and spin $1$
representations respectively on the Wilson lines while HOMFLY and Kauffman
polynomials correspond to $N$-dimensional representation of
$SU(N)$ and $SO(N)$ theories respectively.
We have shown in our earlier work \cite {942,kau} that the link
invariants obtained by putting spin $3/2$
representation on the Wilson lines in $SU(2)$ Chern-Simons theory
do detect chirality of knots upto at least 10 crossings. It appears
going higher in spin, yields more powerful invariants.

It is also known that a class of knots and links, called mutants,
are not distinguished
by Jones, Kauffman/Akutsu-Wadati and HOMFLY polynomials. We shall
demonstrate that even more general Chern-Simons invariants
associated with arbitrary
representations of any compact semi-simple gauge group are not powerful
enough to distinguish mutants. This result follows from the braiding
properties of four-point correlators of the corresponding Wess-Zumino
conformal field theory on $S^2$. In particular,
the commonly referred class of pretzel knots (related to each other by
a sequence of mutations) are not distinguished by
any of these invariants.

To improve the classification scheme, we study the Murakami
$r$-parallel version of braids \cite {mur} in $SU(2)$ Chern-Simons
theory. Instead of the original form of these composite braids, we
shall develop our discussion for a symmetrized, though
equivalent, version.
Representations of the composite braids (symmetrized
version) will be obtained in the basis of conformal blocks of $SU(2)_k$
Wess-Zumino theory. The link invariants constructed from these composite
representations will be referred to as $r$-composite invariants in
contrast to the elementary Chern-Simons invariants.
In fact the $r$-composite invariants for knots are
simply the sum of Chern-Simons invariants associated with the
irreducible representations (allowed by the fusion rules of the
corresponding Wess-Zumino conformal field theory) in the product of
$r$ representations  living on the individual strands constituting the
$r$-composite braid. The composite link invariants also turn
out to be a weighted sum of elementary multicolour
Chern-Simons invariants with the weights given in terms of
the linking number and the colours of the links.
We find that some mutant links are distinguished
by the invariants associated with the composite braids whereas
all mutant knots are not.
The method is general enough to apply to
Chern-Simons theory based on any arbitrary gauge group.

In sec.2, after defining the operation of mutation, we present
a general proof that the field theoretic invariants
do not distinguish
mutants. This is done in the Chern-Simons field theory based
on any arbitrary non-abelian gauge group. In sec.3, we develop
the theory of $r$-composite braids.
We present the representation theory of composite braids
for $r=2$ explicitly in the bases associated with the Hilbert space
of $SU(2)$ Wess-Zumino conformal blocks. The results are then
generalized to $r$-composite braids. In sec.4, the invariants obtained
from the composite braid group representations are presented.
We show that some mutant links are distinguished by the invariants of their
associated composite links. In this context, pretzel links are studied
as an example. In contrast, we demonstrate that the mutant knots
are not detected by the composite knot invariants. Kinoshita-Terasaka
and Conway knots are discussed as an example
of a pair of mutant knots.
We summarize the results in sec. 5.

\section{Mutants and their Chern-Simons invariants}

Let a link $L_1$ be obtained from two rooms
$\cir S$ and $\cir R$ with two strands going in and
two leaving in each of them
as shown in Fig.1(a). The mutant links
are obtained in the following way:
(i) Remove one of the rooms, say $\cir R$ from $L_1$
 and rotate it through $\pi$
about any one of the three orthogonal axes ($\gamma_i$) as shown
in Fig.2. Clearly only two of these rotations are independent:
$\gamma_3~=~\gamma_1 * \gamma_2.$
(ii) Change the orientations of the lines
inside the rotated room $\cirr {${\gamma_i R}$}$ to
match with the fixed orientations of the external
legs of the original
room $\cir {R}$.
(iii) Then, replace this room
back in $L_1$. This yields mutant links
$L_2$ and $L_3$ as shown in Fig.1(b) and (c).

It is known that isotopically distinct mutants have same Jones, HOMFLY and
Akutsu-Wadati/Kauffman invariants. In fact, all the knot polynomials
obtained in the Chern-Simons theory do not distinguish these mutants. To
show this,
observe that the link $L_1$ in $S^3$ can be obtained by gluing a 3-ball
containing room $\cir R$ as shown in Fig.3(a) with another 3-ball
with oppositely oriented boundary $S^2$
containing room $\cir S$ as shown in
Fig.3(d). Similarly, gluing Fig.3(b) and Fig.3(c) with Fig.3(d) will give the
the corresponding mutant links $L_2$ and $L_3$. Further, consider
an $S^3$ with two balls removed from it.
We place four lines, connecting the two boundaries,
in it without any twist and with twists as shown in Fig.4(a), (b)
and (c).
Notice, gluing the three-ball of Fig.3(a) onto the manifold in Fig.4(a)
does not change the three-ball. On the other hand, gluing this
three-ball onto  manifolds in Fig.4(b) and (c) yields the 3-balls shown in
Fig.3(b) and (c) respectively. Thus gluing the manifolds of Fig.4(b) and
(c) onto the manifold of Fig.3(a) is equivalent to introducing rotations
$\gamma_1$ and $\gamma_2$ respectively.

To study the Chern-Simons invariants associated with mutants, we place the
Wilson line operator carrying representation $R$ of gauge group $G$ on the
strands. We are interested in evaluating the Chern-Simons functional
integral over the manifolds of Figs.3 and 4.
These functional integrals represent states
in the Hilbert space
of the 4-point correlator conformal blocks
associated with  4-punctured $S^2$ boundaries \cite{witt}-\cite{kau}.
In particular, the functional integral over the three-manifold
with two boundaries and four untwisted Wilson lines connecting
these boundaries as shown in Fig.4(a) can be represented as \cite{ours}-\cite
{kau}
\begin{equation}
\nu_1~=~
\sum _l~{\ket {\phi_l^{side~(1)}}} {\ket
{\phi_l^{side ~(2)}}}~.
\label {one}
\end{equation}
Here $\ket {\phi_l^{side~(1)}}$ and $\ket {\phi_l^{side~(2)}}$ are the
basis vectors of the two Hilbert spaces associated with two
boundaries of the manifold respectively. Superscript ``side'' on the
basis states indicate that these basis vectors are eigen states of the
braiding generator $b_1$ and $b_3$ introducing half-twists in the first two
or the last two strands:
\begin{equation}
b_1{\ket {\phi_l^{side}}} = b_3 {\ket {\phi_l^{side}}}=
\l_l^{(-)}(R,\bar R)
{\ket {\phi_l^{side}}}~. \label {prop}
\end{equation}
Here the side two strands in Fig.4(a) are antiparallel and carry
representation $R$ and $\bar R$, the index $l$
runs over all the irreducible representations in the fusion rule of
$R\otimes \bar R$
of the corresponding Wess-Zumino model. An equivalent basis
$\ket {\phi_m^{cent}}$ is one where braid generator $b_2$,
which introduces half-twists in the central two strands, is
diagonal:
\begin{equation}
b_2{\ket {\phi_m^{cent}}} =
\l_m^{(+)}(R,R)~\ket {\phi_m^{cent}}~.
\end{equation}
Since this  refers to parallel strands in Fig.4(a) both
carrying representation $R$, the index
$m$ refers to the allowed irreducible representations in the fusion rule
$R \otimes R$ of the corresponding Wess-Zumino model. The eigenvalues
$\l_l^{(-)}(R,\bar R)$ and $\l_m^{(+)}(R,R)$ for antiparallel and
parallel strands are respectively \cite {ours}:
\begin{equation}
\l_l^{(-)}(R,\bar R)~=~(-1)^{\epsilon}~q^{C_l/2}~;~
\l_m^{(+)}(R,R)~=~(-1)^{\epsilon}~q^{2C_R - C_m/2}~, \label {soln}
\end{equation}
where $C_R$, $C_m$ and $C_l$ are the quadratic Casimirs in the representations
$R$, $m$ and $l$ respectively. Depending upon the representation $l$ ($m$)
occuring symmetrically or antisymmetrically in the
tensor product $R\otimes \bar R$ ($R\otimes R$), $\epsilon=\pm1$.
Further  $q~=~\exp {2\pi i/ (k+C_v)}$, where
$C_v$ is the  quadratic Casimir in the adjoint representation and $k$ is
the Chern-Simons coupling.

The two bases are related by
$q$-Racah coefficient of the quantum group $G_q$\cite {alwa,ours}:
\begin{equation}
{\ket {\phi_l^{side}}}~=~\sum_m a_{lm}\left
[\matrix {\bar R&R\cr R&\bar R}\right ]
{\ket {\phi_m^{cent}}}~.\label {prop1}
\end{equation}

The Chern-Simons functional integral $\nu_2$ over the 3-manifold shown
in Fig.4(b) is generated by applying braid
generators $b_1$ and $b_3^{-1}$ on the identity braid of Fig.4(a). Since
$b_1$ and $b_3$ commute and hence are diagonal in the same basis $\ket
{\phi_l^{side}}$ and also have the same eigenvalues (\ref {prop}), the
functional integral $\nu_2$ for manifold of Fig.4(b) is same as that
for the manifold in Fig.4(a):
\begin{equation}
\nu_2~=~
\sum _l~{\ket {\phi_l^{(1)}}} b_1 b_3^{-1} {\ket
{\phi_l^{(2)}}}~=~
\nu_1~. \label {idy}
\end{equation}
Though as a braid Fig.4(a) is isotopically different from Fig.4(b),
the properties of the braid representations in terms
of four-point conformal blocks are responsible
for $\nu_1$ to be equal to $\nu_2$. Such statements will not hold if we
increase the number of Wilson lines in these manifolds.

In order to obtain the action of a $\gamma_2$-mutation on any state,
let us consider the Chern-Simons functional integral $\nu_3$ corresponding
to Fig.4(c). This can be obtained from the state $\nu_1$ representing
the functional integral on the manifold of Fig.4(a) by applying
$b_1b_2b_1b_3b_2b_1$ on it:
\begin{equation}
\nu_3~=~\sum_l {\ket {\phi_l ^{side (1)}}}b_1b_2b_1b_3b_2b_1
{\ket {\phi_l^{side (2)}}}~. \label {stat}
\end{equation}
Now we use the fact that in this Hilbert space
associated with four-punctured $S^2$, $b_1=b_3$ (\ref {prop}).
Further, for an $n$-strand braid on $S^2$, there is an identity
$b_1b_2....b_{n-2}b_{n-1}^2b_{n-2}.....b_2b_1=1$.
This in our case $n=4$, reduces to $b_1b_2b_3^2b_2b_1=1$. This
makes the functional integral $\nu_3$ to be equal to $\nu_1$. Hence,
\begin{equation}
\nu_3=\nu_1=\nu_2~. \label{impo}
\end{equation}

Now let us turn to the Chern-Simons functional integrals for one-boundary
manifolds shown in Fig.3. We shall represent them by vectors
$\ket {\psi_1}$, $\ket {\psi_2}$ and $\ket {\psi_3}$ respectively. These
are related to each other through the functional integrals $\nu_2$ and
$\nu_3$. As stated earlier, gluing the manifold of Fig.4(a) onto that of
Fig.3(a) along an oppositely oriented boundary does not change the
manifold. However, gluing the manifolds of Fig.4(b) and (c) onto that of
Fig.3(a) changes them to the mutants depicted in Fig.3(b) and (c)
respectively. These imply the following relations for the respective
functional integrals:
\begin{equation}
\ket {\psi_1}~=~\nu_1 \ket {\psi_1}~,~~~
\ket {\psi_2}~=~\nu_2 \ket {\psi_1}~,~~~\
\ket {\psi_3}~=~\nu_3 \ket {\psi_1}~.
\end{equation}
where the functional integrals $\nu_1$, $\nu_2$ and $\nu_3$ now refer to
manifolds in Fig.4(a), (b) and (c) with opposite orientations on the two
boundaries. Thus eqn.(\ref{impo}) yields:
\begin{equation}
\ket {\psi_1}~=~
\ket {\psi_2}~=~\ket{\psi_3}~. \label {Res}
\end{equation}

Now the Chern-Simons functional integrals over $S^3$ containing links
$L_1$, $L_2$ and $L_3$ (Fig.1),
$V_R[L_1]$, $V_R[L_2]$ and $V_R[L_3]$,
are given by the products of vector $\bra {\Phi}$ representing
the functional integral over the manifold shown in Fig.3(d)
containing room $\cir {S}$ and  $\ket {\psi_1}$, $\ket
{\psi_2}$ and $\ket {\psi_3}$ representing Figs.3(a), (b) and
(c) respectively:
\begin{equation}
V_R(L_1)= {\norm {\Phi}{\psi_1}},~~~
V_R(L_2)= {\norm {\Phi}{\psi_2}},~~~
V_R(L_3)= {\norm {\Phi}{\psi_3}}.
\end{equation}
Equation(\ref {Res}) then implies
\begin{equation}
V_R[L_1]=V_R[L_2]=V_R[L_3].
\end{equation}

{\it {Thus we have shown that invariants of a link and its mutants are
identical for every
representation $R$
of a compact semi-simple gauge group, placed on all the Wilson
lines constituting the links.}}

As mentioned earlier, the well-known invariants viz., Jones,
HOMFLY and Kauffman
polynomials are obtained from
$SU(2)$, $SU(N)$ and $SO(N)$ Chern-Simons
theories respectively.
Also  Akutsu-Wadati
polynomials \cite {akut} obtained from $N$ state vertex models correspond to
$SU(2)$ with spin $N/2$ representation being placed on the knot/link.
Hence the fact that all these polynomials do not distinguish mutants is
a special case of the above result.

As  an example, we now discuss the class of pretzel links
obtained by stacking $m$ vertical braids with arbitrary number of
half-twists ($a_1,a_2,...a_m$) as shown in Fig.5.
All possible permutations ${\cal P}$ of these vertical braids with different
half-twists ${\cal P}(a_1,a_2,...a_m)$
can be treated as a product of mutations of the
vertical braids in pair. HOMFLY invariants for these links are shown to be
same by Lickorish and Millett \cite {lick}. Our arguments above demonstrate
that even other Chern-Simons invariants with same representation
on the component knots are identical for these links:
$$V_R(L[a_1,a_2,...a_m])~=~V_R(L[{\cal P}(a_1,a_2,...a_m)]).$$

It is appropriate at this point to mention that
some mutant links can be distinguished through multicolour link
invariants Ref.\cite
{ours,kaul} by placing different
representations on the component knots.
These mutant links are those related by mutations
($\gamma_1$, $\gamma_2$ or $\gamma_3$) in rooms containing
strands carrying different representations.

\section{Theory of composite braids}

We saw in the last section that interesting identities of four point conformal
blocks were responsible for the inability of any polynomial invariant
defined from Chern-Simons theory to distinguish the mutant knots and
links. Such identities are not valid for higher point conformal blocks.
Composite braids when placed in a manifold with two $S^2$ boundaries,
result in higher number of punctures on the boundaries.
This raises the hope that one may be able to distinguish mutant knots
and links through invariants constructed from the representations of these
braids.
Theory of $r$-parallel composite braids has been developed by
Murakami \cite {mur}. We shall develop the representation theory of these
composite braids in the conformal block basis.

In Murakami's construction of $r$-parallel
version of the braids, every strand is replaced by
a composite of $r$ strands. The elementary generators $b_i$
($i=1,2...n-1$) of the braid group
${\cal B}_n$ are replaced by $(n-1)$ composite braid
generators $\phi^r(b_i) \in {\cal B}_{rn}$
as shown in Fig.6(a) which depicts
a map from ${\cal B}_n \rightarrow {\cal B}_{rn}$.
In terms of the elementary braid
generators $b_i$, the composite braid generators are given by:
\begin{equation}
\phi^r(b_i)~=~\left (b(ri-r+1,ri-1)\right )^{(-r)}~b(ri,ri+r-1)~
b(ri-1,ri+r-2)~......b(ri-r+1,ri)~.
\label {mura}
\end{equation}
where $b(i,j)= ~b_i b_{i+1}...b_j$.
We shall, however, use a different construction of composite
braids as shown in Fig.6(b) and (c). This one is symmetric in the two composite
legs, each leg is twisted around itself by $\pi$, as against the Murakami
version of Fig.6(a), where only one leg is twisted around itself
by $2\pi$. The
symmetrized $r$-composite braid generators will be denoted by ${\bf B}^{(r)}$.
These generators for parallel composite braids(Fig.6(b))
can be represented in terms of elementary braid generators as:
\newpage
\begin{eqnarray}
{\bf B}^{(r)}(b_i)
{}~=~ b(ri-r+1,ri-1)^{-1} b(ri-r+2,ri-1)^{-1}....
b(ri-1,ri-1)^{-1}\qquad \nonumber\\
\qquad~~b(ri+1,ri+r-1)^{-1} b(ri+2,ri+r-1)^{-1}.... b(ri+r-1,ri-1)^{-1}
\nonumber\\
b(ri,ri+r-1)
b(ri-1,ri+r-2)~... b(ri-r+1,ri)~.
\label {mura2}
\end{eqnarray}
where $b(i,j)=b_ib_{i+1}....b_j$.

The braid generators for the composite braid, like the Murakami
generators \cite {mur}, satisfy the braid group identities:
\begin{eqnarray}
{\bf B}^{(r)}(b_i)~ {\bf B}^{(r)}(b_j)~=~
{\bf B}^{(r)}(b_j)~ {\bf B}^{(r)}(b_i)~,
 ~~~for~~ |i-j| > 1 \label{brai}\\
{\bf B}^{(r)}(b_i)~ {\bf B}^{(r)}(b_{i+1})
{}~{\bf B}^{(r)}(b_i)~=~ {\bf B}^{(r)}(b_{i+1})
{}~{\bf B}^{(r)}(b_i)~ {\bf B}^{(r)}(b_{i+1})~.
\end{eqnarray}

The closures of composite braids give knots and links. The
closure respects the invariance
under Markov moves (Fig.7):
\begin{eqnarray}
closure(AB)=closure(BA)~, \\
closure(A~{\bf B}^{(r)}(b_n^{\pm 1}))= closure(A)~. \label {clos}
\end{eqnarray}
where $A$ and $B$ are elements of composite braid group
${\cal B}_n$.

The above composite braid formalism can be generalised for
antiparallel strands also.
This will enable us to obtain any knot or link
by platting of braids \cite {kaul,kau}. Such a braid
is drawn in Fig.6(c). A corresponding representation in terms of the
elementary braid generators can also be written down.

We shall now develop explicit representations for the composite braid
generators ${\bf B}^{(r)}(b_i)$ in the $SU(2)$ conformal block basis.
For simplicity, the discussion will be presented for
composite braids made of two strands, $r=2$. Generalizations to other
values of $r$ is straightforward.

For $r=2$, the generators for composite parallel braids can be written as:
\begin{equation}
{\bf B}^{(2)}(b_i)~=~b_{2i-1}^{-1}~b_{2i+1}^{-1}
b(2i,2i+1)~
b(2i-1,2i)~.
\label {symm}
\end{equation}
Similarly, the generators for the antiparallel
composite braids are:
\begin{equation}
{\bf B}^{(2)}(b_i)~=~b_{2i-1}~b_{2i+1}
b(2i,2i+1)~
b(2i-1,2i)~.
\label {symm1}
\end{equation}

The representations of braid generators ${\bf B}^{(2)}(b_i)$ can
be obtained in terms of the representations
of the elementary braiding generators $b_i$ and the duality matrix
(\ref {prop1}) relating the eigen bases of $b_i$ and $b_{i+1}$
in the $SU(2)$ Chern-Simons framework. Let us consider a four strand braid
carrying a spin $j$ representation. Associated composite braid with
$r=2$ will have eight elementary strands.
We start with the identity operator $\chi_1$
in a manifold with two $S^2$ boundaries as shown in
Fig.8(a) and act
on this by ${\bf B}^{(2)}(b_2)$ to obtain
a composite braid $\chi_2$ as shown in Fig.8(b).
The Chern-Simons functional integrals over mainfolds $\chi_1$ and $\chi_2$
can be expanded in terms of  convenient conformal block bases
associated with Wess-Zumino theory on the two eight-punctured $S^2$
boundaries in the same manner as in Ref.\cite {kaul}. We shall show that
the composite braid ${\bf B}^{(2)}(b_2)$ is diagonal in the conformal block
bases $\ket {\phi}$ of $SU(2)$ Wess-Zumino theory drawn in Fig.9(a).
In this figure, at
every trivalent point, the various spins obey the fusion rules of the
$SU(2)$ Wess-Zumino conformal field theory.
Thus we write
the functional integrals over manifolds $\chi_1$ and $\chi_2$ as:
\begin{eqnarray}
{\chi_1}=\sum_{l_i,m_j}
\ket {\phi_{l_1,(l_2,l_3,n_1),l_4}^{(1)}}
\ket {\phi_{l_1,(l_2,l_3,n_1),l_4}^{(2)}}~,~~~~~~~~~~~~~~~\label {cald}\\
\chi_2=
{\bf B}^{(2)}(b_2)
\chi_1=\sum_{l_i,m_j}
{\ket {\phi_{l_1,(l_2,l_3,n_1),l_4}^{(1)}}}
{\bf B}^{(2)}(b_2)
{\ket {\phi_{l_1,(l_2,l_3,n_1)l_4}^{(2)}}}~,
\end{eqnarray}
where the superscript $(1)$ and $(2)$ on the basis vectors corresponds to
the two $S^2$ boundaries.

Writing the right-handed parallel composite braid generator
explicitly in terms of the elementary
braid generators, $b_3^{-1}b_5^{-1}b_4b_5b_3b_4$,its
eigenvalues  as indicated in Appendix, turn out to be:
\begin{eqnarray}
{\bf B}^{(2)}(b_2)
{\ket {\phi_{l_1,(l_2,l_3,n_1),l_4}}}
{}~&=&~ {\tilde \l}_{n_1}^{(+)}(l_2,l_3)
{\ket {\phi_{l_1,(l_3,l_2,n_1),l_4}}}~,\nonumber\\
{\tilde \l}_{n_1}^{(+)}(l_2,l_3)~&=&~
(-1)^{n_1}q^{C_{l_2}+C_{l_3}-C_{n_1}/2}~.\label {res1}
\end{eqnarray}
Similarly, the eigen basis for the commuting composite generators
${\bf B}^{(2)}(b_1)$ and ${\bf B}^{(2)}(b_3)$ can be shown
to be that associated with the eight point
conformal blocks of Fig.9(b) denoted as
${\ket {{\hat \phi}_{(l_1,l_2,m_1),(l_3,l_4,m_1)}}}$ with
eigenvalues ${\tilde \l}_{m_1}^{(+)}(l_1,l_2)$ and ${\tilde
\l}_{m_1}^{(+)}(l_3,l_4)$ for parallel composite braids, respectively.

Following Appendix, it is straightforward to work out the eigenvalues
for the righthanded antiparallel composite braid operator of Fig.6(c).
These eigenvalues for ${\bf B}^{(2)}(b_2)$ are  found to be
\begin{equation}
{\tilde \l}_{n_1}^{(-)}(l_2,l_3)=(-1)^{n_1}
q^{C_{n_1}/2}~.\label {res2}
\end{equation}
The absence of $l_2$ and $l_3$ in the expression on the right-hand side,
in contrast to that for parallel braids (\ref {res1})
is consistent with first Reidemeister move. The eigen basis for this operator
is the same as for the corresponding parallel composite braid above.
Notice that these
eigenvalues do not explicitly depend on the spin $j$
of the external eight lines of the conformal bases. These depend only
on the spins on the internal lines.
The above discussion has been developed for symmetrized version of the
composite braids (Fig.6(b),(c)). The eigenvalues of the braid matrix here
have a symmetric form. For the original composite braids of Murakami
(Fig.6(a)), these eigenvalues are different. For parallel $(+)$
and antiparallel  $(-)$ composite braids, these turn out to be
${\hat \l}_{n_1}^{(\pm )}(l_2,l_3)~=~(-1)^{l_2\pm l_3}q^{{C_{l_3}-C_{l_2} \over
2}}{\tilde
\l}_{n_1}^{(\pm )}(l_2,l_3)$.

The eigen bases of odd-indexed generators ${\bf B}^{(2)}(b_1)$,
${\bf B}^{(2)}(b_3)$ and even-indexed generator
${\bf B}^{(2)}(b_2)$ are related by the four-point
duality matrix relating the internal lines with spins $l_1,l_2,n_1,l_3,l_4$
in Fig.9(a) and spins $l_1,l_2,m_1,l_3,l_4$ in Fig.9(b):
\begin{equation}
{\norm {\hat {\phi}_{(l_1,l_2,m_1),(l_3,l_4,m_1)}}
{\phi_{l_1,(l_2,l_3,n_1),l_4}}}~=~
a_{m_1 n_1}\left [ \matrix{l_1&l_2\cr l_3& l_4}\right]~.
\end{equation}
Notice again, like the eigenvalues of composite braids
above, it is the duality
matrix relating only the internal spins that appears here.

We have completely determined the theory of composite braid for
the $r=2$ case. Generalisation to $r$-composite braids
with arbitrary number $(n)$ of strands can be easily done. The relevant
conformal block
bases (Fig.10(a),(b)) are the extensions of
the conformal blocks (Fig.9(a),(b)).
The  $l_i$ ($i\in [1,n]$) are the allowed representations in the tensor
product of the $r$ spin $j$ representations of the
$SU(2)$ Wess-Zumino model. The composite strands made up of $r$
individual strands, all carrying spin $j$, have been represented by thick
external lines as shown in Fig.10(c).

The odd-indexed composite braid generator
${\bf B}^{(r)}(b_{2i+1})$
is diagonal in the
$SU(2)$ conformal block basis $\ket {\hat {\phi}}$ of Fig.10(a):
\newpage
\begin{eqnarray}
{\bf B}^{(r)}(b_{2i+1})
{\ket {\hat {\phi} \left[(l_1,l_2,m_1),
s_0,(l_3,l_4,m_2),s_1,...s_{i-1},(l_{2i+1},l_{2i+2},m_{i+1}),..
\right ]}}~=~~~~~~~~~~~~~\nonumber \\
{\tilde \lambda _{m_{i+1}}(l_{2i+1},l_{2i+2})}
{\ket {\hat {\phi}\left[(l_1,l_2,m_1),
s_0,(l_3,l_4,m_2),s_1,...s_{i-1},(l_{2i+2},l_{2i+1},m_{i+1}),s_i,
....\right ]}}~,\label {eigen}
\end{eqnarray}
where the eigenvalue for parallel $(+)$
and antiparallel $(-)$ braid generators introducing righthanded half-twists
are:
\begin{eqnarray}
\tilde \l_{m_{i+1}}^{(+)}(l_{2i+1},l_{2i+2})~&=&~(-1)^{2l_{2i+1}-m_{i+1}}
q^{C_{l_{2i+1}}+C_{l_{2i+2}}-C_{m_{i+1}}/2}~,\label {Rep}\\
\tilde \l_{m_{i+1}}^{(-)}(l_{2i+1},l_{2i+2})~&=&~
(-1)^{m_{i+1}}q^{C_{m_{i+1}}/2}~~~~~~~~~~~~~~~.\label {Rep1}
\end{eqnarray}
The eigenvalues are symmetric under interchange $l_{2i+1}
\leftrightarrow
l_{2i+2}$. (Notice that the $l_i$s are either all integer or all half-integers
and hence $(-1)^{2l_i}=(-1)^{2l_j}$). The half-twist operator ${\bf
B}^{(r)}(l_{2i+1})$, in addition to interchanging the internal labels
$l_{2i+1}$ and $l_{2i+2}$ of the conformal block also reverses the order of
the external $r$-legs of each composite strand (thick line)
connected to these two internal lines.

The eigen basis $\ket {\phi}$ for the even indexed composite generator
${\bf B}^{(r)}(b_{2i})$  are diagonal in the basis of Fig.10(b):
\begin{eqnarray}
{\bf B}^{(r)}(b_{2i})
{\ket {\phi \left[l_1,(l_2,l_3,n_1),
r_1,(l_4,l_5,n_2),r_2,...r_{i-1},(l_{2i},l_{2i+1},n_i),r_i,..
..\right ]}}~=~~~~~~~~~~~~~~~\nonumber \\
{\tilde \lambda _{n_i}(l_{2i},l_{2i+1})}
{\ket {\phi \left[l_1,(l_2,l_3,n_1),
r_1,(l_4,l_5,n_2),r_2,...r_{i-1},(l_{2i},l_{2i+1},n_i),r_i,..
..\right ]}} ~,\label {eigen1}
\end{eqnarray}
with eigenvalues given by eqn.(\ref {Rep}) and (\ref {Rep1}) for parallel
and antiparallel righthanded twists.

Using methods of Ref.\cite {kaul}, the
generalised transformation  matrix relating the
two bases in eqns.(\ref {eigen}, \ref {eigen1}) (Fig.10(a) and (b)) can
be seen to
be:
\begin{equation}
\norm  {\hat {\phi}}{\phi}~=~
\prod _i a_{r_i m_{i+1}} \left[ \matrix {s_{i-1} & l_{2i+1}\cr l_{2i+2}
& s_i}\right]
{}~\prod _i a_{s_{i-1} n_i} \left[ \matrix {r_{i-1} & l_{2i}\cr
l_{2i+1}
& r_i}\right]~,~~~~~~~~~\label {res3}
\end{equation}
where $r_0=l_1$ and $s_0=m_1$.

{\it Equations (\ref {eigen} - \ref {res3})
constitute the composite braid representations.}

Thus it is clear that only the irreducible representation in the product of
spin $j$ carried by the individual elementary strands
in a $r$-composite strand appear
in the representation theory of the composite braids, {\it i.e.,} in the braid
eigenvalues (\ref {Rep},\ref {Rep1}) and generalised duality matrix (\ref
{res3}).

This  completes our discussion of representation theory of composite
braid in the $SU(2)$ Chern-Simons framework.
The formalism developed here can be
extended to any compact semi-simple group. Further, we have placed the same
set of $r$-representations in each composite strand. We could as well
place different sets of $r$-representations on the different composite
strands. Theory of such multicoloured composite braids can also be
developed in a similar fashion.

\section{Mutants and composite braids}

After the above discussion of the representation theory of composite braids,
we now take up the question whether
mutant knots/links can be distinguished by
invariants obtained from these new representations of the
braid group.
{}From sec.1, we know that
it was the result in eqn.(\ref {impo}) asserting the equality of
Chern-Simons functional integrals
associated with Fig.4(a), (b) and (c),
that was responsible for mutants to have same elementary
Chern-Simons invariants. This result depended on the fact that the
functional integrals over the manifolds of Fig.4(a), (b) and (c) could be
written as states in the product Hilbert spaces associated with
four punctured $S^2$ boundaries. For $r$-composite braids, the
corresponding Hilbert spaces are those associated with boundaries
with $4r$ punctures. This opens up the possibility that the invariants
obtained through composite braid representations above may distinguish
mutant knots/links. The manifolds corresponding to Fig.4(a), (b) and (c)
are drawn with composite braids ($r=2$)
in Fig.8(a), 11(a) and (b)
respectively. We shall denote the Chern-Simons functional integrals
over these manifolds as $N_1$, $N_2$ and $N_3$ respectively.
We write down the explicit representations for them in terms of the
$r$-composite braid representations. The functional integral
$N_1$ associated with Fig.8(a) for $r$-composite braid is:
\begin{equation}
N_1~=~ \sum_{(l),m_1}
\ket {\hat {\phi}_{(l_1,l_2,m_1),(l_3,l_4,m_1)}^{(1)}}
\ket {\hat {\phi}_{(l_1,l_2,m_1),(l_3,l_4,m_1)}^{(2)}}~,~~~~~ \label{idy5}
\end{equation}
where the conformal blocks $\ket {\hat {\phi } }$
correspond to Fig.10(a) and the superscript (1) and (2) refer to the two
boundaries. The Chern-Simons functional integral $N_1$ here written in
terms of basis in which the odd-indexed composite braid generator are
diagonal is the same as $\chi_1$ in eqn.(\ref {cald}) where we have
represented it in a basis in which the even index generators are diagonal.
The functional integral $N_2$ associated with Fig.11(a) can be written as:
\begin{eqnarray}
N_2~&=&~ \sum_{(l),m_1}
\ket {\hat {\phi}_{(l_1,l_2,m_1),(l_3,l_4,m_1)}^{(1)}}
{\bf B}^{(r)}(b_1)
({\bf B}^{(r)}(b_3))^{-1}
\ket {\hat {\phi}_{(l_1,l_2,m_1),(l_3,l_4,m_1)}^{(2)}}\nonumber \\
{}~&=&~{\tilde \lambda} _{m_1}^{(-)}(l_1,l_2)
({\tilde \lambda} _{m_1}^{(-)}(l_3,l_4))^{-1}
\ket {\hat {\phi}_{(l_1,l_2,m_1),(l_3,l_4,m_1)}^{(1)}}
{\ket {\hat {\phi}_{(l_2,l_1,m_1),(l_4,l_3,m_1)}^{(2)}}}~,\nonumber \\
{}~&=&~\ket {\hat {\phi}_{(l_1,l_2,m_1),(l_3,l_4,m_1)}^{(1)}}
{\ket {\hat {\phi}_{(l_2,l_1,m_1),(l_4,l_3,m_1)}^{(2)}}}~.~~~~~~~~~
{}~~~~~~~~~~~~~~~~~~~~\label{nidy}
\end{eqnarray}
Here eqn.(\ref {Rep1}) has been used.
In a similar manner, the composite state
$N_3$ associated with the $r$-composite braid of Fig.11(b) is:
\begin{eqnarray}
N_3~=~ \sum_{(l),m_1}
\ket {\hat {\phi}_{(l_1,l_2,m_1),(l_3,l_4,m_1)}^{(1)}}
{\bf B}^{(r)}(b_2) {\bf B}^{(r)}(b_1)
{\bf B}^{(r)}(b_3) {\bf B}^{(r)}(b_2)]\nonumber
\\
{}~[{\bf B}^{(r)}(b_3)]^{-1} [{\bf B}^{(r)}(b_1)]
^{-1}
\ket {\hat {\phi}_{(l_1,l_2,m_1),(l_3,l_4,m_1)}^{(2)}}~, \label {ques1}
\end{eqnarray}
The result of the composite braid ${\bf B}^{(r)}(b_2) {\bf B}^{(r)}(b_1)
{\bf B}^{(r)}(b_3) {\bf B}^{(r)}(b_2) [{\bf B}^{(r)}(b_3)]^{-1} [{\bf
B}^{(r)}(b_1)]^{-1}$, is simply to interchange the internal spins in the
basis vector $\ket {\hat {\phi}_{(l_1,l_2,m_1),(l_3,l_4,m_1)}}$ to
$\ket {\hat {\phi}_{(l_4,l_3,m_1),(l_2,l_1,m_1)}}$. This is precisely
how a $\gamma_2$-mutation would act on such a representation. This can also
be seen explicitly by substituting the composite braid representation
into eqn.(\ref {ques1}). After some algebra and use of identities
(A.5, A.6) from the Appendix, the following result is obtained:
\begin{equation}
N_3~=~ \sum_{(l),m_1}
\ket {\hat {\phi}_{(l_1,l_2,m_1),(l_3,l_4,m_1)}^{(1)}}
\ket {\hat {\phi}_{(l_4,l_3,m_1),(l_2,l_1,m_1)}^{(2)}}~. \label {nidy1}
\end{equation}

Thus clearly from eqns.(\ref {idy5}, \ref {nidy} and \ref {nidy1}),
we find :
\begin{equation}
N_1~\neq N_2~\neq
N_3 ~.\label {Res1}
\end{equation}
This result is encouraging and thus may allow us to distinguish mutants.
In particular, let us consider an example of a pair of mutant links
belonging to pretzel class $L_1=L[3,2,2,3]$ and $L_2=L[3,2,3,2]$ of Fig.5.
These two links are related by a $\gamma_1$-mutation.
The link $L_1$ can be thought of as
obtained by gluing two copies of the $3$-ball drawn in Fig.12(a) onto
each other along oppositely oriented boundaries. On the other hand,
mutant link
$L_2$ is obtained by gluing the ball of Fig.12(a) onto that in Fig.12(b).
We shall study these links with the orientations as specified in these
figures. We replace the strands in Fig.12(a) and (b) by composites with
$r$ elementary strands. The various braids are also changed to symmetrised
composite braids. The corresponding Chern-Simons functional integral
associated with these manifolds (Fig.12(a) and (b)) containing
$r$-composite braids will be represented by vectors
$\ket {\Psi_1}$ and $\ket {\Psi_2}$ respectively.
Following methods of Ref.\cite {kaul}, these can be expressed
in terms of conformal blocks associated with $S^2$ boundary containing
$4r$ punctures. A straightforward calculation yields (after using
identities A.5, A.6, and A.7):
\begin{eqnarray}
\ket {\Psi_1}~=~ \sum_{l_1,l_2,r} f(l_1,l_2,r)
\ket {{\hat {\phi}_{(l_1,l_2,r),(l_2,l_1,r)}}}~,\\
\ket {\Psi_2}~=~ \sum_{l_1,l_2,r} f(l_1,l_2,r)
\ket {{\hat {\phi}_{(l_2,l_1,r),(l_1,l_2,r)}}}~,
\end{eqnarray}
where
$$f(l_1,l_2,r)~=~\sum_{s_1,s_2}
(-1)^{2l_1+r+s_1}{[2s_2+1]\sqrt {[2r+1]} \over [2l_2+1]}
q^{C_{s_2}-8C_{l_1}-2C_{l_2}+{3 \over 2}C_{s_1}}
\left (a_{l_2 s_1}\left [\matrix {l_1 & s_2\cr r & l_1} \right ] \right)^2~.
$$
The square bracket represents the $q$-numbers:
$[x]~=~(q^{x/2}-q^{-x/2})/(q^{1/2}-q^{-1/2})$.
These states are related to each other by
$$ \ket {\Psi_2}~=~N_2 \ket {\Psi_1}$$
where $N_2$ (Fig.11(a)) here has oppositely oriented two $S^2$ boundaries.
It is also clear that the coefficients in
the state $\ket {\Psi_1}$ is not symmetric under
the interchange $l_1 \leftrightarrow l_2$. This implies:
$\ket {\Psi_1} \neq \ket {\Psi_2}$ which is the reflection of the
fact $N_1 \neq N_2 $.

The composite invariant for the oriented link $L_1$ is the product of the
state $\ket {\Psi_1}$ with its dual; for
oriented link $L_2$, it is the inner
product of $\ket {\Psi_1}$ with $\ket {\Psi_2}$:
\begin{eqnarray}
V_j[L_1]~=~\langle {\Psi_1}| {\Psi_1}\rangle ~=~
\sum_{l_1,l_2,r} \left (f(l_1,l_2,r)\right )^2~,\\
V_j[L_2]~=~\langle {\Psi_2}| {\Psi_1}\rangle~=~
\sum_{l_1,l_2,r} f(l_1,l_2,r) f(l_2,l_1,r)~.
\end{eqnarray}
Again, clearly these are not equal: $V_j[L_1] \neq V_j[L_2]$. In
particular, for $j=1/2$ on every individual strand in the $r$-composite
braid (Jones Polynomial), a straightforward computation yields:
\begin{equation}
V_{1/2}[L_1] ~-~V_{1/2}[L_2]~=~q^{-4}(1+q+q^{-1})(X -1)^2
\end{equation}
where
$X~=~(q^{-11}-q^{-10}-q^{-9}+q^{-8}-q^{-7}+q^{-5}+q^{-2})$.

Next let us take up an example of a pair of mutant knots, namely the
famous 11 crossing Kinoshita-Terasaka
knot and its mutant known as the Conway knot as shown in
Fig.13(a) and (b). The Kinoshita-Terasaka knot $K_1$ can be thought of as
obtained by gluing three-ball of Fig.14(b) onto that of Fig.14(c) along
oppositely
oriented boundaries. The Conway knot $K_2$ in contrast, is obtained by gluing
the three-ball in Fig.14(a) onto the three-ball in Fig.14(c).
The state $\ket {\Psi_3}$ associated with Fig.14(a)
can be readily evaluated by the method of Ref(\cite {kaul}):
\begin{equation}
\ket {\Psi_3}~=~
\sum_{l_1,l_2,r} A(l_1,l_2,r)
\ket {\hat {\phi}_{(l_1,l_1,r),(l_2,l_2,r)}}~,
\end{equation}
where $A(l_1,l_2,r)$ is:
\begin{eqnarray}
A(l_1,l_2,r)~=\sum_{m,n}~[2l_2+1]\sqrt {[2l_2+1][2l_1+1]}a_{0 m}\left
[\matrix {l_1&l_1\cr l_2 & l_2 }\right ]
\left (\tilde {\l}_m^{(-)}(l_1,l_2) \right )^2
a_{0n}\left [\matrix {l_2&l_2\cr l_2 & l_2} \right ]\nonumber \\
\left (\tilde {\l}_n^{(+)}(l_2,l_2) \right )^{-3}
a_{r m}\left [\matrix {l_1&l_1\cr l_2 & l_2} \right ]
a_{r n}\left [\matrix {l_2&l_2\cr l_2 & l_2} \right ]
a_{l_2 0}\left [\matrix {r&l_2\cr l_2 & r} \right ]~.
\end{eqnarray}
The state $\ket {\Psi_4}$ associated with
Fig.14(b) is related to state $\ket {\Psi_3}$ through $N_3$:
\begin{equation}
\ket {\Psi_4}~=~N_3 \ket {\Psi_3}~=~
\sum_{l_1,l_2,r} A(l_1,l_2,r)
\ket {\hat {\phi}_{(l_2,l_2,r),(l_1,l_1,r)}}~.
\end{equation}
Similar calculation for the state $\ket {\Psi_5}$ associated with
Fig.14(c), yields:
\begin{equation}
\ket {\Psi_5}~=~\sum_{(l_1,l_2,r} B(l_1,l_2,r) \ket
{\hat {\phi}_{(l_1,l_2,r),(l_1,l_2,r)}}~,
\end{equation}
where $B(l_1,l_2,r)$ is given by:
\begin{eqnarray}
B(l_1,l_2,r)~&=&~\sum_{(n),(m),(y),(m')} \sqrt {[2l_1+1][2l_2+1]}~ [2l_2+1]
a_{0m_1} \left[\matrix {l_1& l_1\cr l_2 & l_2}\right ]\nonumber \\
{}~&~&a_{0m_2} \left[\matrix {l_2& l_2\cr l_2 & l_2}\right ]
{\tilde \l}_{m_1}^{(+)}(l_1,l_2)
{}~[{\tilde \l}_{m_2}^{(-)}(l_2,l_2)]^{-1}
a_{n_1m_1} \left[\matrix {l_1& l_2\cr l_1 & l_2}\right ]\nonumber\\
{}~&~&a_{n_2m_2} \left[\matrix {l_2& l_2\cr l_2 & l_2}\right ]
a_{l_2 y_1} \left[\matrix {n_1& l_1\cr l_2 & n_2}\right ]
({\tilde \l}_{y_1}^{(+)}(l_1,l_2))^{-1}
a_{y_2y_1} \left[\matrix {n_1& l_2\cr l_1 & n_2}\right ]\nonumber\\
{}~&~&a_{n_1m_1'} \left[\matrix {l_1& l_2\cr l_2 & y_2}\right ]
a_{n_2m_2'} \left[\matrix {y_2& l_1\cr l_2 & l_2}\right ]
({\tilde \l}_{m_1'}^{(+)}(l_2,l_2))^2
[{\tilde \l}_{m_2'}^{(-)}(l_1,l_2)]^{-1}\nonumber\\
{}~&~&a_{r m_1'} \left[\matrix {l_1& l_2\cr l_2 & y_2}\right ]
a_{r m_2'} \left[\matrix {y_2& l_2\cr l_1 & l_2}\right ]
a_{y_20} \left[\matrix {r& l_2\cr l_2 & r}\right ]~.
\end{eqnarray}
The composite invariant for the Kinoshita-Terasaka knot $K_1$
is given by the inner product of $\ket {\Psi_4}$ and
$\ket {\Psi_5}$ and that for the Conway knot $K_2$ by the inner product of
$\ket {\Psi_3}$ and $\ket {\Psi_5}$:
\begin{equation}
V_R[K_1]~=~\langle \Psi_5| \Psi_4 \rangle~=~\sum_{l,r} A(l,l,r)
B(l,l,r)~=~\langle \Psi_5| \Psi_3 \rangle~=~V_R[K_2]~.
\end{equation}
Thus composite invariants do not distinguish these mutants. In fact,
this is true for all mutant knots. This can be seen as follows:
Consider a 3-ball containing a room $\cir {S}$ with
four external legs marked as
$A,B,C,D$ as shown in the Fig.15. There are three ways in which
these four legs can be connected to each other  inside the room:

Case (i)
$A$ is connected to $B$, $C$ is connected to $D$
inside the room like in Fig.12(a) and (b).
Then  a general state associated with the Fig.15 for this
case will be
\begin{equation}
{\ket {\tilde {\Psi}_1}}~=~\sum_{(l),r} F(l_1,l_2,r)
\ket {\hat {\phi}_{(l_1,l_2,r),(l_2,l_1,r)}}~,
\end{equation}
where $F(l_1,l_2,r)$ is  a function which depends on the
room contained in the three-ball.
In general, this function need not be invariant
under interchange of $l_1$ with $l_2$.

Case (ii)
$A$ is connected to $C$, $B$ is connected to $D$
inside the room as in Fig.14(a) and (b). Then the general state
will be
\begin{equation}
{\ket {\tilde {\Psi}_2}}~=~\sum_{(l),r} G(l_1,l_2,r)
\ket {\hat {\phi}_{(l_1,l_1,r),(l_2,l_2,r)}}~.
\end{equation}

Case (iii)
$A$ is connected to $D$ and $C$ is connected to $B$ inside the room
as in Fig.14(c). In this case the
state will be
\begin{equation}
{\ket {\tilde {\Psi}_3}}~=~\sum_{(l),r} H(l_1,l_2,r)
\ket {\hat {\phi}_{(l_1,l_2,r),(l_1,l_2,r)}}
\end{equation}

Observe that, we could get knots by gluing two  3-balls such
that (a) one satisfies case (i) and the other case (ii), (b) one satisfies
case (i) and the other case (iii), and (c) one satisfies case (ii) and the
other case (iii). Then the composite invariants for three knots $K_1$, $K_2$
and $K_3$ so constructed are:
\begin{eqnarray}
V_R[K_1] =\sum_{l,r}F(l,l,r)G(l,l,r)~;\nonumber \\
V_R[K_2] = \sum_{l,r}F(l,l,r)H(l,l,r)~;\nonumber\\
V_R[K_3] = \sum_{l,r}G(l,l,r)H(l,l,r).
\end{eqnarray}
These knot invariants obtain  contribution  only from the subspace
spanned by the basis vectors
$\ket {\hat {\phi}_{(l,l,r),(l,l,r)}}$. Also, from eqns.(\ref {idy5}-\ref
{nidy1}), the
state responsible for mutation will behave like an identity state in this
subspace. {\it {i.e}}.,
\begin{equation}
N_1|_{l_i=l}~=~N_2|_{l_i=l}
{}~=~N_3|_{l_i=l}~.
\end{equation}
Hence, the mutants of any knot cannot be distinguished even by composite
invariants.

Clearly, the composite braid eigenvalues (\ref {Rep}, \ref {Rep1}) and
transformation matrices
(\ref {res3})
in this invariant subspace
are the same as the eigenvalues (A.3,A.4) and duality matrices for
elementary braids carrying spin $l$ representations.
Hence the $r$-composite invariant for a given knot
is simply the sum over the Chern-Simons invariants for the irreducible
representations in the product of $r$ spin $j$
representations. This is true not only
for $SU(2)$ but, by a straightforward generalisation of above
arguments, also for other compact semi-simple Lie groups. Thus this
result can be stated as:

{\it {For a general gauge group $G$, the $r$-composite knot invariant
associated
with representation $R$ of $G$ is the sum of elementary Chern-Simons invariants
for all irreducible representations in the product of $r$
representations $R\otimes R \otimes....R$ as
allowed by the fusion rules of corresponding Wess-Zumino conformal
field theory.}}

Since elementary Chern-Simons
invariants for any gauge group do not distinguish mutant knots, it is
clear, not even composite invariants  can do so.
In particular, composite version of the HOMFLY polynomial also cannot
distinguish mutant knots in contrast to results in\cite {mur}.

Now let us discuss the case of mutant links. Links can be
obtained by gluing each of the states corresponding to case (i), (ii)
and (iii) onto its dual or mutated dual state. Notice that the
state $\ket {\tilde {\Psi}_1}$ is invariant under $\gamma_2$
mutation, and the state $\ket {\tilde {\Psi}_2}$ is invariant under
$\gamma_1$ mutation. Then mutant link (if isotopically distinct) which
can be constructed from $\ket {\tilde {\Psi}_1}$ ($\ket {\tilde
{\Psi}_2}$) and its $\gamma_2$ ($\gamma_1$) mutated dual state
cannot be distinguished by composite invariants. On the
other hand mutant link obtained from $\ket {\tilde {\Psi}_1}$ ($\ket {\tilde
{\Psi}_2}$) and its $\gamma_1$ ($\gamma_2$) mutated dual state
can be distinguished (if isotopically distinct) by
composite invariants. The example of the  class of pretzel
links (Fig.5) discussed above is of this type: they can be constructed
by the products of $\ket {\tilde {\Psi}_1}$ with its dual
and its $\gamma_1$ mutated dual respectively.

In fact the composite link invariants are related to elementary
multicoloured link invariants. This follows from the fact that
the eigenvalues (\ref {Rep}, \ref {Rep1}) of the composite braid
and elementary multicoloured braid generators (A.3,A.4) are related as:
\begin{equation}
\tilde {\l}_{n_1}^{(\pm )}(l_2,l_3)~=~(-)^{l_2-l_3}q^{|C_{l_2}-C_{l_3}|\over 2}
\l_{n_1}^{(\pm )}(l_2,l_3)~.\label {quot}
\end{equation}
Thus the composite invariant $V_j^{comp}[L]$ for a link $L$ can be written
in terms of the elementary multicolour link invariants
$V_{l_1,l_2,...,l_m}[L]$,
where $l_1,l_2,...,l_m$ are the spins placed on the $m$ component
knots $K_1,K_2,...,K_m$ of the link $L$:
\begin{equation}
V_j^{comp}[L] ~=~ \sum_{l_1,l_2,..,l_m} q^{\sum_{s,t} {\it {lk}}(s,t)
{}~|C_{l_s}- C_{l_t}|}
{}~V_{l_1,l_2,...l_m}[L]~, \label {link}
\end{equation}
The $l_i$ span the spins of irreducible representations in the product of $r$
spin $j$ representations
carried by the individual strands in the composite
braid. Here ${\it {lk}}(s,t)$ is the linking number for the $(K_s,K_t)$
pair of component knots in $L$. Notice that the $q$-independent
phase factors $(-)^{l_2-l_3}$ in the eigenvalues (\ref {quot}) disappear
in eqn.(\ref {link}) because
$l_2-l_3$ is always an integer and there are even number of crossings between
any
pair of linking component knots. The result in eqn.(\ref {link}) generalises to
invariants
obtained from any arbitrary compact semi-simple group.

Though the composite braiding eigenvalues for the asymmetric and
symmetric composite braids (Fig.6) are different, the composite invariants
for knots and links constructed from these two types of composite
braids are the same.

\section{Conclusions}

In this paper, we have presented a proof that the isotopically
inequivalent mutant knots and links cannot be distinguished by
the elementary Chern-Simons invariants associated with the representations of
any gauge group $G$. The proof involves the fact that independent
mutations $\gamma_1$ and $\gamma_2$ on the Chern-Simons functional
integral over a three-ball with four punctured $S^2$ boundary do not
change it (eqn.\ref {Res}). This follows from specific braiding properties
of four-point conformal blocks of the corresponding Wess-Zumino conformal
field theory.

In order to explore the possibility that composite braid representations may be
of help to distinguish mutant knots and links, we have  developed the
representation theory of such braids made up of $r$ strands within the
framework of Chern-Simons field theory. The
composite representations of generators of
the $n$-braids can be given in the basis of $nr$-point conformal blocks of
the corresponding Wess-Zumino theory. These
do not depend explicitly on the spins
placed on individual strands in the $r$-composite braid, but depend only
on the spins on the internal lines of the corresponding conformal
blocks (Fig.11(a) and (b)).

The $r$-composite invariant for a knot carrying
representation $R$ of the group $G$ is given by the sum of elementary
Chern-Simons
invariants for the knot associated with the irreducible representations
in the product of $r$ representations, $R\otimes R\otimes ....R$,
allowed by the fusion rules of the
corresponding Wess-Zumino conformal field theory. Thus, it is clear that
mutant knots cannot be distinguished by such composite invariants.
On the other hand, we have argued that the composite invariants for a link
can be written as a weighted sum of multicoloured elementary invariants (\ref
{link}).
Therefore,
some mutant links do have distinct composite invariants.
These are the links related by a mutation of a four-leg room carrying
strands from two distinct component
knots. Specific examples
discussed are the class of pretzel links. We have clearly
demonstrated that the $r$-composite invariants associated with any
representation of $SU(2)$ are indeed different. In particular, spin $1/2$
composite invariants, which corresponds to  Jones polynomials, distinguish
these mutant links. However, links related by a mutation of a four-leg
room carrying strands from the same component knot cannot be distinguished
by composite invariants.

Since higher spin elementary Chern-Simons invariants tend to
distinguish chirality of knots \cite {942,kau}, it is worth
pointing out that the composite knot invariants
will also tend to do so. However, there are exceptions.
For example, the sixteen crossing
knots (Fig.16(a) and (b)) discussed in ref\cite {kauff} are respectively chiral
and achiral. These are also related by a mutation.
Hence the chirality of knot in Fig.16(a) cannot be detected by
any elementary Chern-Simons invariant, nor by any composite invariant.
\vskip.5cm

\noindent
\appendix{\large {\bf Appendix}}
\vskip.5cm

In this Appendix, following methods of ref.\cite{kaul} we outline the
derivation of eigenvalues of the composite
braid
operator for right-handed half-twists in the parallel composite ($r=2$)
strands.
The operation of the
symmetrized composite braid operator
${\bf B}^{(2)}(b_2)$  (Fig.8(b)) on the eight-point conformal block of Fig.9(a)
is as follows:
$${\bf B}^{(2)}(b_2)
{\ket {\phi_{l_1,(l_2,l_3,n_1),l_4}}}
{}~=~b_3^{-1}b_5^{-1}b_4b_5b_3b_4
{\ket {\phi_{l_1,(l_2,l_3,n_1),l_4}}}~.\eqno (A.1)
$$
The basis chosen is not diagonal in the even
indexed elementary generator
$b_4$. Using appropriate duality matrices (\ref {prop1}), we go to its
diagonal basis to operate $b_4$. Then, again we have to
go back to the same odd-indexed basis to operate $b_3$ on it.  This
leads to
\begin{eqnarray}
b_3~b_4{\ket {\phi_{l_1,(l_2,l_3,n_1),l_4}}}
{}~=~ \sum_{[(r),(l'),(m),n_1]}~
a_{m_1n_1}\left [\matrix{l_1&l_2\cr l_3 &l_4}\right]
a_{r_2l_2}\left [\matrix{l_1&j\cr j &
m_1}\right] a_{r_3l_3}\left[\matrix {m_1&j\cr j &l_4}\right ]\nonumber\\
a_{m_1,p_1}\left[\matrix {r_2 & j\cr j &r_3}\right]
\l_{p_1}^{(+)}(j,j) a_{m_2 p_1}\left[\matrix{r_2 &j\cr j&r_3}\right]
a_{r_2l_2'}\left[\matrix {l_1&j\cr j &m_2}\right ]\nonumber
\end{eqnarray}
$$~~~~~~~~~~~~~~~~~~~~~~~~~~~~~~~~~a_{r_3l_3'}\left[\matrix {m_2&j\cr j
&l_4}\right ]
a_{m_2n_2}\left[\matrix {l_1&l_2'\cr l_3' &l_4}\right ]
\l_{l_2'}^{(+)}(j,j)
{\ket {\phi_{l_1,(l_2',l_3',n_2),l_4}}}~,\eqno (A.2)
$$
where $\l_n^{(\pm)}(j_1,j_2)$ are the eigenvalues of elementary braiding
matrices for righthanded half-twist in parallelly $(+)$ and antiparallelly
$(-)$ oriented strands, carrying spins $j_1$ and $j_2$,
given by (\cite {kaul}):
$$\l_n^{(+)}(j_1,j_2)~=~(-)^{j_1+j_2-n}q^{C_{j_1}+C_{j_2}-~{|C_{j_1}-C_{j_2}|
\over 2}~- C_n/2}~,\eqno (A.3)$$
$$\l_n^{(-)}(j_1,j_2)~=~(-)^{|j_1-j_2|-n}q^{-~{|C_{j_1}-C_{j_2}|
\over 2}~+ C_n/2}~,\eqno (A.4)
$$
and the four-point $SU(2)$ duality matrices $a_{ij}\left
[\matrix {l_1 & l_2\cr l_3&l_4}\right ]$ are explicitly given in Appendix
I of Ref.\cite {kaul}.

Similarly the generators $b_3^{-1}b_5^{-1}b_4b_5$ can be applied.
The following orthogonality condition and identities \cite {kaul} satisfied
by the $q$-Racah coefficients enables us to simplify the
expression:
$$
\sum_{m_2} a_{r_1 m_2}\left[ \matrix {m_1 & l_1\cr l_2& m_3}\right
]a_{r_2 m_2}\left[ \matrix {m_1 & l_1\cr l_2& m_3}\right ]~=~
\delta_{r_1 r_2}~,~~~~~~~~~~~\eqno (A.5)$$
$$\sum_{s_1} a_{r_1 s_1} \left[ \matrix{ p_1&j_1 \cr j_2&p_2 } \right]
a_{r_2 s_1} \left[ \matrix{ p_1&j_2 \cr j_1&p_2 } \right]~(-1)^{s_1}
q^{-C_{s_1/2}}~~~~~~~~~~~~~~~~~~~~~~~~~~~~~~~~~ $$
$$~~~~~~~~~~~~~~~~~~~~~~~~~=~(-1)^{-r_1-r_2+p_1+p_3+j_1+j_2} a_{r_1 r_2}\left
[\matrix {j_2 &
p_2\cr j_1& p_1}\right ] q^{(C_{r_1}+C_{r_2}-C_{j_1}-C_{j_2}-
C_{p_1}-C_{p_2})/2}~,\eqno (A.6)$$
$$\sum_{p_1}a_{l_1 p_1} \left[ \matrix{ r_1&m_1 \cr j_1&j_2 } \right]
a_{r_1 r_2} \left[ \matrix{ p_2&j_3 \cr p_1&j_2 } \right]
a_{p_1 s_1} \left[ \matrix{ r_2&j_3 \cr m_1&j_1 }
\right]\qquad\qquad~~~~~~~~~~~~~~~~$$
$$~~~~~~~~~~=~a_{l_1 r_2} \left[ \matrix{ p_2&s_1 \cr j_1&j_2 } \right]
a_{r_1 s_1} \left[ \matrix{ p_2&j_3 \cr m_1&l_1 }
\right]~.~~~~~~~~~~~~~~~~~~~~~~~\eqno (A.7)$$
Finally, we arrive at the simplified result for two-strand parallel composite
braids:
$$
{\bf B}^{(2)}(b_2)
{\ket {\phi_{l_1,(l_2,l_3,n_1),l_4}}}
{}~=~(-1)^{n_1} q^{C_{l_2}+C_{l_3}-C_{n_1}/2}
{\ket {\phi_{l_1,(l_3,l_2,n_1),l_4}}}
\eqno (A.8)
$$

Starting with its representation in terms of elementary braid
generators as given in eqn.(\ref {symm1}) a similar calculation can be done for
the eigen values of an antiparallel composite braid operator of Fig. 6(c).

\vskip1cm
\flushleft{\bf Figure captions:}\\
\begin{itemize}
\item[Fig.1] Mutant links.
\item[Fig.2] Mutations ($\gamma_1, \gamma_2, \gamma_3$) of room $\cir
{R}$.
\item[Fig.3] Three-balls, each with four punctured $S^2$ boundary.
\item[Fig.4] Diagrammatic representations of the Chern-Simons functional
integrals
for  3-manifolds with two boundaries: (a) $\nu_1$, (b) $\nu_2$ and (c)
$\nu_3$.
\item[Fig.5]Pretzel link $L[a_1,a_2....a_m]$.
\item[Fig.6]$r$-composite braids: (a) Murakami braid $\phi^{(r)}(b_i)$,
(b) symmetrized parallel braid
${\bf B}^{(r)}(b_i)$
and (c) symmetrized antiparallel braid ${\bf B}^{(r)}(b_i)$.
\item[Fig.7] Markov moves for composite braids.
\item[Fig.8] Diagrammatic representations of functional integrals (a) $\chi_1$
and (b) $\chi_2$
\item[Fig.9] Eight point conformal blocks bases
(a) $\ket {\phi}$ and (b) $\ket {\hat {\phi}}$.
\item[~Fig.10]Eigen bases (a) ${\ket {\phi}}$ for odd indexed $r$-composite
braid generator,
(b) $\ket {\hat {\phi}}$ for even indexed $r$-composite braid generator
and (c) explicit representation of the composite thick line in terms of
$r$ lines each carrying spin $j$.
\item[Fig.11] Diagrammatic representations of Chern-Simons functional
integrals over two-boundary manifolds $N_2$ and $N_3$.
\item[Fig.12] Diagrammatic representations of the states (a) $\ket
{\Psi_1}$ and (b) $\ket {\Psi_2}$.
\item[Fig.13] Eleven crossing mutants: (a)Kinoshita-Terasaka knot $K_1$ and
(b) Conway knot $K_2$.
\item[Fig.14] Diagrammatic representations of the functional integrals
over three-balls: (a) $\ket {\Psi_3}$,
(b) $\ket {\Psi_4}$ and (c) $\bra {\Psi_5}$
\item[Fig.15] A three-ball containing room $\cir S$ with marked points
$A,B,C,D$
on the $S^2$ boundary.
\item[Fig.16] A pair of mutant knots with 16 crossings.
\end{itemize}
\end{document}